\documentclass{article}
\usepackage[T1]{fontenc}
\usepackage[latin1]{inputenc}
\usepackage{graphics,epsfig}
\usepackage{amssymb}
\usepackage{pslatex}

\begin{document}

\title{Non-locality: what are the odds?}

\author{A. F. Kracklauer \footnote{\tt kracklau@fossi.uni-weimar.de}}

\date{}

\maketitle

\begin{abstract}As part of a challenge to critics of Bell's analysis of
the EPR argument, framed in the form of a bet, R. D. Gill formulated 
criteria to assure that all non-locality is precluded from
simulation-algorithms used to test Bell's theorem. This is achieved in
part by parceling out the subroutines for the source and both detectors
to three separate computers. He argues that, in light of Bell's theorem,
following these criteria absolutely precludes mimicking EPR-B
correlations as computed with Quantum Mechanics and observed in
experiments. Herein, nevertheless, we describe just such an local
algorithm, fully faithful to his criteria, that yields results mimicking
exactly quantum correlations. We observe that our simulation-algorithm
succeeds by altering an implicit assumption made by Bell to the
equivalent effect that the source of EPR pairs is a single Poisson
process followed by deterministic detection. Instead we assume the
converse, namely that the source is deterministic but detection involves
\emph{multiple}, independent Poisson processes, one at each detector
with an intensity given by Malus' Law. Finally, we speculate on some
consequences this might have for quantum computing algorithms.
\end{abstract}

\section{State of Play}

Richard D. Gill, a statistician, on the basis that probability and
statistics have much to do with quantum mechanics (QM), and by his
own declaration fascinated by the exotica of foundations disputes,
has called for increased attention to these matters by his profession.\cite{1}
To this purpose, he himself has published studies that translate some
of the current notions oft seen in the foundations of QM into the
jargon and style of his discipline.\cite{2} So far Gill's results
have tended to support the conventional wisdom, the majority viewpoint
held by, for example, several who have done experiments that they
credit even with extending the mystical facets of QM right into the realm
of practical applications; e.g., quantum computing, teleportation
and the like.

In addition, Gill has criticized the views of those  who are
distressed by the implications of modern foundations' orthodoxy of
QM, specifically and particularly that the non-locality promised by
John Bell violates the order of cause and effect.   \footnote{A
renowned proponent of Bell's view told this writer, on the grounds
that humankind probably is simply incapable of grasping all truth,
that he was not distressed by this inversion. If, however,
\emph{Science} is that human enterprise, the point of which is to use
the tools of logic (Mathematics) and observation as available to
understand the world, then such an orientation is in a distinctly
different category.} Some of their efforts are based on a point first
seen, to this writer's best information, by Jaynes: Bell misapplied
Bayes's formula.\cite{3}

An independent study in this vein initiated by Accardi, as an
``anti-Bellist,'' led to a local computer simulation of EPR-B
experiments that he claimed exhibit the so-called non-local statistics
of QM.\cite{4} Accardi's simulation fascinated Gill, who found the
tactic meritorious but the results so far unconvincing, and so
challenged Accardi to find a protocol that is beyond dispute. Convinced
that it can't be done, and to enliven the matter, Gill offered a 3000
Euro reward, essentially a bet, to Accardi should he
succeed.\cite{1,5} 

The present writer is also one long sceptical of arguments that non-locality is
logically well founded.\cite{6} He holds the view that some essential feature
is being overlooked, much as von Neumann overlooked crucial contrary physics
details in formulating his ``theorem'' to the same final effect as Bell's.
This writer's work led to a series of arguments indicating that Bell's
argumentation must contain a flaw.\cite{7,7.1} Eventually this culminated in a
study in which all the results from generic experiments credited with verifying
Bell's results were calculated using only principles of classical Physics,
without reference to anything intimating that non-locality was in play.\cite{8}
In spite of a vigorous e-mail defense, these calculations failed to impress
Gill, in part, apparently, for lack of clarity on how non-locality was
precluded. Gill continued to insist that no protocol satisfying certain
constraints that he formulated to enforce locality, could be envisioned.

It is the purpose of this note to present an extention of these
calculations in the form of a simulation algorithm or protocol
conforming with Gill's criteria. In the following section the
conditions set out by Gill will be presented; thereafter, the
simulation we propose will be described. Finally, we delineate the
technical details that make our simulation possible.

\section{Gill's desiderata}

The purpose of Gill's protocol is to preclude absolutely any structure that
could covertly exploit non-locality.  This is achieved by parceling out
subroutines simulating the source and the two detector stations to three
separate computers connected by wires conceptually equipped with diodes that
allow only one way signaling. In this way, one can be sure that there is no
feedback in the logic that mimics non-local interaction. This writer holds
these specifications as a useful contribution to the discussion of this
matter because, for lack of a unique definition of ``non-locality,''
alternates that are, e.g., actually only indirect statistical
congruences, have clouded the matter. 

Gill's specifications are as follows,\cite{5} we quote:

\begin{quote}
1. Computer \textbf{O}, which we call the \emph{source,} sends information
to computers \textbf{X} and \textbf{Y}, the \emph{measurement stations.}
It can be anything. It can be random (previously stored outcomes of
actual random experiments) or pseudo-random or deterministic. It can
depend in an arbitrary way on the results of past trials (see item
5). Without loss of generality it can be considered to be the same---send
to each computer, both its own message and the message for the other.

2. Computers \textbf{A} and \textbf{B,} which we call randomizers,
send each a \emph{measurement-setting-label,} namely a $1$ or a $2$,
to computers \textbf{X} and \textbf{Y}. Actually, I will generate
the labels to simulate independent fair coin tosses (I might even
use the outcomes of real fair coin tosses, done secretly in advance
and saved on my computers' hard disks.

3. Computers \textbf{X} and \textbf{Y} each output $\pm 1$, computed
in whatever way {[}an opponent{]} likes from the available information
at each measurement station. He has all the possibilities mentioned
under item 1. What each of these two computers do not have, is the
measurement-setting-label which was delivered to the other. Denote
the outcomes $x^{(n)}$ and $y^{(n)}$.

4. Computers \textbf{A} and \textbf{B} each output the measurement-setting-label
which they they had previously sent to \textbf{X} and \textbf{Y.}
Denote these labels $a^{(n)}$ and $b^{(n)}$. An independent referee
will confirm that these are identical to the labels given to {[}an
opponent{]} in item 2.

5. Computers \textbf{X, O} and \textbf{Y} may communicate with one
another in any way they like. In particular, all past setting labels
are available to all locations. As far as I am concerned, {[}an opponent{]}
may even alter the computer programs or memories of his machines.
\end{quote}
At this point Gill proceeds to delineate how the data is to be analyzed.
What he calls for is the total of the number of times the outputs
are equal, i.e., $N_{ab}^{=}:=\#[x^{(n)}=y^{(n)}]$, as well as the
number of times they are unequal, $N_{ab}^{\neq }$ , and the total
number of trials, $N_{ab}$. With these numbers, the correlation,
$\kappa $, for each setting pair, $ab$, is then:\begin{equation}
\kappa _{ab}=\frac{N_{ab}^{=}-N_{ab}^{\neq }}{N_{ab}}.\label{corr}\end{equation}
 These correlations, in turn, are used to compute the CHSH contrast:\begin{equation}
S=k_{12}+k_{11}+k_{21}-k_{22},\label{CHSH}\end{equation}
 which is to be tested for violation of Bell's limit, $|S|\leq 2$,
as is well known.

\section{A counter-protocol}

Computer \textbf{O} is implemented as an equal, flat, random selection
of one of two possible signal pairs, one comprised of a vertically
polarized pulse to the left, say, and a horizontally polarized pulse
to the right; the second signal exchanges the polarizations. 

After the source pulse pair is selected, the measurements to be simulated at
computers \textbf{X} and \textbf{Y} are selected by independent computers
\textbf{A} and \textbf{B}. This they do for each run or  each pulse pair
individually by randomly specifying which of the two orientation angles for
each side and iteration is to be used; i.e., they select  $\theta _{l}$ and
$\theta _{r}$. 

The measurement stations \textbf{X} and \textbf{Y} are simulated by
models of polarizers for which the axis of the left (right) one is the
angle $\theta _{l\, (r)}$; each polarizer feeds a photodetector. These
photodetectors are taken to adhere to Malus' Law, that is, they produce
photoelectron streams for which the intensity is proportional to the
incoming field intensity, and the arrival time of the photoelectrons is
a random variable described by a Poisson process. In so far as these
photodetectors are independent, each Poisson process is uncorrelated
with respect to the other. In conformity with Bell's assumptions and the
experiments, we take it that the source power and pulse duration are so
low that, within a time-window, one and only one photoelectron will be
generated if the polarization of the signal and the axis of the
polarizer preceeding the detector are parallel.  Thus, when they are not
parallel, the count rate is reduced in proportion to the usual Malus'
factor. This assumption is structurally parallel to assuming single
photon states in QM.  It is unrealistic to the extent that, even for the
parallel regime, a true Poisson distribution would lead to some trials
with no hits, which don't contribute to the photoelectron count
statistics, and some trials with two or more hits, which are so seldom
they effectively don't contribute.

As a matter of detail, the photoelectron generation process is modeled
by comparing a random number with the intensity of the field entering the
polarizer filter. If it is less, it is taken that a
photoelectron was generated; if greater, none was generated. In other
words, $N_{lv}$ say, (where for $N_{sm}$, $s$ indicates the setting of
$\theta _{s}$ as chosen by \textbf{A} or \textbf{B}; and $m$ indicates
pulse mode from the source: vertical or horizontal) is increased by $1$,
when the random number $\leq \cos ^{2}(\theta _{a}-\phi _{l})$.  Also,
$N_{rh}$ is increased by $1$ when another, independent random number
$\leq \cos ^{2}(\theta _{b}-\phi _{r})$. In each case $\phi _{j=r,l}=0\,
(\textrm{vertical),}\: \pi \textrm{/2}\, \textrm{(horizontal) }$is the
polarization axis of the pulse sent from the source. The random input at
the detector simulates background signals and detector noise.

The final step of the simulation is simply to register the simulated
`creation' of photoelectrons and to compute the correlations.

This algorithm fully satisfies Gill's stipulations as presented in Section 2
above to preclude non-locality. All process steps can be implemented on
separate computers such that the information flow is from \textbf{A},
\textbf{B} and \textbf{O} to \textbf{X} and \textbf{Y}. There is no
connection between \textbf{X} and \textbf{Y}.  At this point we note,
however, that Gill's analysis of EPR experiments is not faithful to the
relevant physics.  It fails to take the quadratic relationship between the
source intensity and the subsequent measured current density, i.e., Malus'
Law, into account.

Thus, in our simulation, data acquisition and analysis proceed a bit
differently. We take it, that \emph{currents} are measured, that the total
relative intensity between channels is given purely by geometrical
considerations according to Malus' Law; i.e., according to $\cos ^{2}(\theta
_{r}-\theta _{l})$ for like events, and $\sin ^{2}(\theta _{r}-\theta _{l})$
for unlike events. This is essentially equivalent to requiring internal self
consistency with respect to detection physics. That is, if the axis of one
polarizer is parallel to the axis of the source, $\theta _{r}=0$, say, then
it is quite obvious that the relative intensity as measured by
photodetectors on the output of the other polarizer must follow Malus' Law
by virtue of the photocurrent generation mechanism. Further, since this
geometric fact must be independent of the choice of coordinate system, it
follows that a transformation of angular coordinates can be effected using
\begin{eqnarray} \cos (\theta _{r}-\theta _{l}) & = & \cos (\theta
_{r})\cos (\theta _{l})+\sin (\theta _{r})\sin (\theta _{l}),\nonumber
\\ \sin (\theta _{r}-\theta _{l}) & = & \sin (\theta _{r})\cos (\theta
_{l})+\cos (\theta _{r})\sin (\theta _{l});\label{trig} 
\end{eqnarray}
just trigonometric identities. Note that although equations (\ref{trig}) are
not in general factorable (sometimes said to be a criteria for
`non-locality'), all the information required is available on-the-spot at
both sides independently, i.e., locally. Thus, in modeling the ``coincidence
circuitry'' we use the fact that the individual terms on the right side of
(\ref{trig}) are, by virtue of photodetector physics, proportional to the
square root of the number of counts in the channel, for example:  
\begin{eqnarray} 
\cos (\theta _{l}-\phi_{h}) =\cos(\theta_{l} -0) =\cos(\theta_{l})  =\lim 
_{N\rightarrow \infty }  \sqrt{N_{hl}/N},\nonumber \\ 
\cos (\theta _{l}-\phi_{v}) =\cos(\theta_{l} -\pi/2) =\sin(\theta_{l}) =\lim
_{N\rightarrow \infty } \sqrt{N_{vl}/N};\label{key} 
\end{eqnarray}
where $N$ is the total number of signals intercepted by the considered
photodetector per regime. (Since there are four detector- and two 
source-regimes, for even distributions \(N \rightarrow T/8\) where \(T\) is
the total number of pairs used in the simulation. Unlike experiments, this
number is trivially available in simulations.)  In other words, the relative
frequency of coincidences is determined after-the-fact as a function of the
intensity of the photocurrents. No communication between right and left
sides is involved in determining these currents; what correlation there is,
is there on account of the equality of the amplitudes (and therefore
\emph{intensities}) at the source of the signals making up the singlet
state. Of course, in doing a simulation, provision must be made to resolve
sign ambiguities; doing so, however, also does not violate locality as the
required information is all available on-the-spot. In sum, instead of Eq.
(\ref{corr}), we use:
\begin{equation} 
\kappa ^{*}(\theta _{r},\,\theta _{l})=\cos ^{2}(\theta
_{r}-\theta _{l})-\sin ^{2}(\theta _{r}-\theta _{l}),\label{korr}
\end{equation} 
as required by Malus' Law, but in which the terms are
expressed using (\ref{trig}) and (\ref{key}) to get the result for each
individual `photoelectron' locally and independently.

To see intuitively how this algorithm works, it is perhaps advantageous to
consider the individual steps in reverse order. The final step is the
calculation of the CHSH contrast. This requires calculating $\kappa
^{*}(\theta _{r},\,\theta _{l})$ for the four combinations of polarizer
settings according to Eq. (\ref{korr}).  Now, each factor  $\kappa ^{*}$, a
coincidence correlation, according to Eq. (\ref{korr}) is expressed as the
intensity of coincidences in like channels minus the intensity in unlike
channels, all according to conventional formulas. The individual terms in
$\kappa ^{*}$, however, involve information from each side, so it would
appear that each requires in effect instantaneous communication between the
sides for evaluation. However, each term can be expanded using Eqs.
(\ref{trig}) in which the individual factors, e.g., $\cos(\theta_l)$,  can
be computed at the detection event without regard to information from other
events. These individual factors, in turn, are provided by Eq. (\ref{key}),
which again is just Malus' Law. As the simulation is repeated, and more and
more pairs of pulses are generated, the running ratios $N_{ms}/N$ converge
to provide ever improving estimates of the individual factors needed for Eq.
(\ref{trig}). As described above, the individual $N_{ms}$ are completely
determined by the \emph{delayed} input from the source, the local polarizer
setting, and a random input at the detector. \emph{Only the after-the-fact
calculation of $\kappa^*$ mixes information from both sides.} Thus, there is
no non-locality involved --- contrary to Bell's claim that a model without
it does not exist.

An example of the results from the simulation are presented on Fig.1.

\begin{figure}[th]
\centerline{\epsfig{file=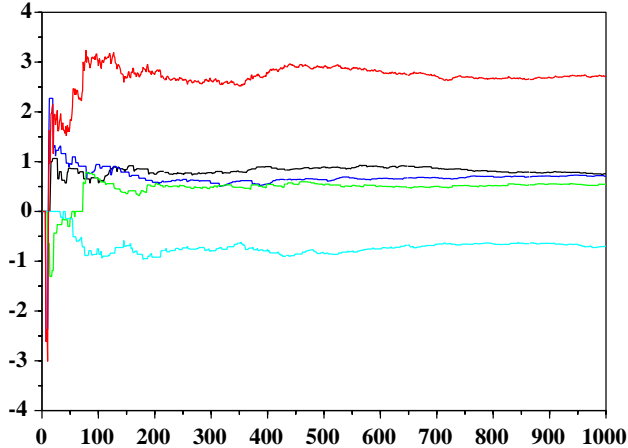,width=10cm}}
\vspace*{8pt}
\caption{An example of a simulation of an EPR-B
experiment.  The top curve is the CHSH contrast; the others are the individual
correlation coefficients, all plotted as a function of the number of
trials.}
\end{figure}

The top curve is the CHSH contrast; the lower four curves are the
individual correlations for the four combinations of polarizer settings.
The statistics stabilize after circa 700 trials and exhibit clear
violation of the Bell limit of $2$  by virtually exactly
the amount calculated for the singlet state and observing angles $\theta _{l}=0,\; \pi /4$
and $\theta _{r}=\pm \pi /8$ using Bell inequalities: $2\sqrt{2}$.
Experimental verification can be found in \cite{9}. That these statistics
can be found in a sequence of individual events, each of which is
calculated without recourse to non-locality, constitutes a direct,
unambiguous counterexample to Bell-type ``theorems.''

As an aside, we note that what is described here is a \emph{simulation,}
not an experiment. As the algorithm runs on a digital computer, in
fact it is a realization of an EPR-B setup using \emph{information,}
which accords with some variations of current analysis of the issue.

\section{Reconciliation}

How does this protocol work? How is it possible that now, after
generations have examined and reexamined seemingly everything
surrounding EPR-B correlations and von Neumann's and Bell's theorems, a
simple protocol can be found that delivers the offending statistics
without non-locality?

The answer to these questions is in part, that so far the implied challenge
has been so defined, that the option we used in constructing the simulation
was precluded. The basic assumption constraining the orthodox approach is
that the source emits ``ready-made'' pairs of photons. At an abstract
statistical level, this is tantamount to assuming that there is just a
single Poisson process at the source, for which the time of conception of
each pair is a random variable with a Poisson distribution, followed by
deterministic detection. That is, in the terminology of QM, the probability
of the random creation of a photon pair is proportional to the modulus of a
solution to Schr\"odinger's equation, followed by detection with a
``quantum'' efficiency of \(\backsimeq 100\)\% of converting photons to
photoelectrons \emph{without further stochastic input} (except at the
polarizer, see below).  

For our protocol we reject these notions. Instead, we take it that the
source is emitting pulses of classical, continuous radiation in both
directions, which are finally registered in photodetectors that are the
scene of independent Poisson detection processes proportional to the square
of the source intensity. Although uncorrelated in terms of arrival times of
elicited photoelectrons, the two processes, by virtue of the  symmetry of
the source (i.e., the source pulses have equal power and duration),
nevertheless have correlated \emph{intensities}.  For the simulation,
consistent with classical physics, polarizers are taken to be  variable,
linear attenuators, which are, therefore, deterministic.  In  `quantum'
imagery, polarizers are considered biased stochastic absorbers, i.e., not
deterministic.  The latter notion  is not relevant to the issue of Bell
inequalities, however, because they are meant to be limits imposed by local
realism and are non-quantum from the start. In the end, the patterns seen in
the correlations are due simply to the various Malus' factors attenuating
the equal energy source pulses.

Independently, underneath all the confused aspects of the dispute initiated
by EPR, there is an inviolable mathematical truth, which has many forms. It
is that both correlated and uncorrelated equal length dichotomic sequences
with values $\pm 1$, tautologically satisfy Bell Inequalities. Being an
ineluctable mathematical truth, it is also often mistaken for a physical
indispensability.

There is, however, an intervening complication. Consider four dichotomic
sequences comprised of $\pm 1$'s and length $N$: $a,$ $a',$
$b$ and $b'$. Now compose the following two quantities
$a_{i}b_{i}+a_{i}b_{i}'=a_{i}(b_{i}+b_{i}')$ and
$a_{i}'b_{i}-a_{i}'b_{i}'=a_{i}'(b_{i}-b_{i}')$, sum them over $i$, divide
by $N$, and take absolute values before adding together to get:
\begin{eqnarray}
|<ab>+<ab'>|+|<a'b>-<a'b'>|\leq <|a||b + b'|>+<|a||b - b'|>\label{BI}
\end{eqnarray}
The right side equals $2$, so this equation is in fact a Bell inequality.
This derivation demonstrates that this Bell inequality is simply an
arithmetic tautology. Thus, certain dichotomic sequences comprised of
$\pm 1$'s,  identically satisfy Bell inequalities.\cite{10}

The fact is, however, that real data can not comply with this inequality; it
just does not fit the conditions of the derivation of Eq. (\ref{BI}).  This
is a result of the fact that two of the sequences on the left side are
\emph{counterfactual} statements, i.e., what \emph{would have been measured}
if the setting \emph{had been} otherwise.  In real experiments, such data
can never be obtained.  Moreover, real data can not be rearranged to fit Eq.
(\ref{BI}) either.  Suppose to start, that the second term is rearranged so
that the factor sequence \(a(2)\) where the argument \(2\) indicates that it
is from the second term, is rearranged to match as closely as possible
\(a(1)\).  (For ever longer samples, this becomes ever more precise.)  Let
the rearranged version be denoted \(\tilde{a}(2)\).  Then the second term
becomes: \(a(2)b'(2)\Rightarrow \tilde{a}(2)\tilde{b}'(2)\cong
a(1)\tilde{b}'(2)\).  A similar rearrangement on the third and fourth terms
converts the right side of Eq. (\ref{BI}) to:
\begin{eqnarray}
<|a(1)||b(1)+\tilde{b}'(2)|>+<|\tilde{a}(4)||\tilde{b}(3)-\tilde{b}'(2)|>, \label{BII}
\end{eqnarray}
from which it is obvious that unless $b(1)\cong \tilde{b}(3)$ by virtue of
the structure of the setup, that Eq. (\ref{BI}) is not germane; in other
words, data from real experiments can not conform to the conditions of
derivation of Bell inequalities. Variations of this argument hold for all
derivations of Bell inequalities.\cite{11} 

\section{Secondary arguments}

Note also that the results of our simulation are in full accord with
Jaynes' criticism of Bell's arguments, to wit: Bell, by ascribing
the correlations to the hidden variables, effectively misused Bayes'
formula for conditional probabilities with respect to the overt variables.
The result is that he encoded statistical independence instead of
non-locality, so that the resulting inequalities are valid only for
uncorrelated sequences. In ref \cite{5}, Gill endeavors to respond
to this observation with the argument that Jaynes ``refuses to
admit'' that:\begin{equation}
P(a|b,\: \lambda )=P(a|\lambda ),\label{hidden}\end{equation}
 where $P(c|d)$ is the \emph{conditional} probability for $c$ occurring
if $d$ has occurred, $a$ and $b$ represent filar marks on a measurement
device and $\lambda $, represents a ``hidden variable'' specifying
the quality whose magnitude influences the magnitude registered as
$a$ and $b$.\cite{5} Of course, if one is measuring an assortment
of objects (nails, say), each engraved with its length ($\lambda $),
then we can dispense with the comparison to the filar marks on a ruler
($a$) in favor of using the $\lambda $. But when the $\lambda $
are not knowable, not to mention uncontrollable, the measured objects
must be characterized by the result of comparison with the ruler,
i.e., the $a$'s. As this is exactly the situation with respect to
EPR and Bell's arguments, all statistical characteristics, including
correlations, of an ensemble must be specified also in terms of results
from measurements in the form of  comparisons with the filar marks,
the only variables actually available. EPR's purpose was to discern
the possible existence of such hidden variables exclusively in terms
of their effects on quantities expressed in terms of overt variables;
all evidence must appear, therefore, in the overt variables. That
is, when Jaynes rejects (\ref{hidden}), he is just faithful to reality
if $\lambda $ remains unaccessible.

The conventional understanding also overlooks the fact that the space of
polarization is fundamentally not quantum mechanical in nature.\cite{5}
This is so in the first instance, because its variables, the two states
of polarization, are not Hamiltonian conjugates, and therefore do not
suffer Heisenberg uncertainty (HU), and do not involve Planck's
constant. Gill rejected this argument on the basis of the opinion that
``quantum mechanics is as much about incompatible observables as
Planck's constant.'' In essence this view implies that \emph{all}
commutivity somehow involves QM, which is manifestly false. Non
commutivity can arise for different reasons, for example, non parallel
Lorentz boosts do not commute, and this obviously has nothing to do with
QM. In the case at hand, non commutivity of polarization vectors, or
Stokes' operators, arises only when the $k$-vector common to both
polarization states rotates in space, which brings in the geometric fact
that the generators of rotations on the sphere do not commute. This has
nothing specifically to do with QM, EPR or Bell inequalities, just
geometry.

\section{Conclusions}

A consequence of the existence of a local-realistic simulation of
EPR-B correlations is that it shows that such correlations do not
arise because of the structure of QM. This could have significant
practical consequences. In particular, to the extent that algorithms,
usually considered to depend on exploiting quantum entanglement, for
example, can be seen not to be dependant on intrinsically quantum
phenomena. Thus, these algorithms should also not be restricted to
realization at the atomic level, where QM reigns. In so far as it
is much more practical to fabricate and operate devices at a larger
scale, this may be a very promising development for the exploitation
of what is (from this viewpoint, inaccurately) denoted as quantum computing.

In the end, does this simulation put Gill in debt? Whatever the call, it is
clear that his considerations, like that of many others, was based on some
mathematical facts regarding dichotomic sequences, comprised of both hits
and non-hits, and not on the actual details of done experiments. These
mathematical facts concerning dichotomic sequences can also be simulated by
doing the data analysis as  envisioned by Gill, and indeed this seems
not to lead to a violation of Bell inequalities no matter how the parameters
are adjusted. In short, what is shown here is, that EPR-B correlations are a
direct manifestation of Malus' Law, not mystical gibberish or QM.

\end{document}